\RequirePackage{ifpdf}
\documentclass{JHEP3}
\usepackage{subfig,amsmath,amssymb}
\usepackage{graphicx}
\usepackage{slashed}
\DeclareMathOperator{\Tr}{Tr}
\newcommand{\LL}{\hat{L}}
\newcommand{\be}{\begin{equation}}
\newcommand{\ee}{\end{equation}}

\title{Mass anomalous dimension of Adjoint QCD at large $N$ from twisted 
volume reduction}

\author{ Margarita Garc\'{\i}a P\'erez$^{a}$, Antonio Gonz\'alez-Arroyo
$^{a,b}$, Liam Keegan$^{c}$ \quad\quad\quad\quad\quad\quad and Masanori Okawa$^{d,e}$ \\
  $^a$ Instituto de F\'{\i}sica Te\'orica UAM-CSIC, Nicol\'as Cabrera 13-15, \\
  Universidad Aut\'onoma de Madrid, E-28049--Madrid, Spain \\
  $^b$ Departamento de F\'{\i}sica Te\'orica, C-XI \\
       Universidad Aut\'onoma de Madrid, E-28049--Madrid, Spain \\
  $^c$ PH-TH, CERN, CH-1211 Geneva 23, Switzerland\\
  $^d$ Graduate School of Science, Hiroshima University,\\
Higashi-Hiroshima, Hiroshima 739-8526, Japan \\
  $^e$ Core of Research for the Energetic Universe, Hiroshima University,\\
Higashi-Hiroshima, Hiroshima 739-8526, Japan \\

E-mail: \email{margarita.garcia@uam.es, antonio.gonzalez-arroyo@uam.es, liam.keegan@cern.ch, okawa@sci.hiroshima-u.ac.jp}
 }

\abstract{In this work we consider the $SU(N)$ gauge theory with two Dirac fermions in the adjoint 
representation, in the limit of large $N$. In this limit the
infinite-volume physics of this model can be studied by means of the
corresponding  twisted reduced model defined on a single site lattice. 
Making use of this strategy we study the reduced model for various
values of $N$ up to 289.  By analyzing the eigenvalue  distribution of the
adjoint Dirac operator we test the conformality of the theory and
extract the corresponding mass anomalous dimension.}


\preprint{CERN-PH-TH-2015-138\\ IFT-UAM/CSIC-15-060\\ FTUAM-15-17 \\HUPD-1501}

\date{\today}

\begin{document}

\section{Introduction}

The $SU(2)$ gauge theory with two adjoint Dirac fermions,
known as Minimal Walking Technicolor
(MWT)~\cite{Sannino:2004qp,Luty:2004ye}, has been the subject of many
lattice studies, all of which have found it to be a conformal theory
with a fairly small mass anomalous dimension $\gamma_*$~\cite{Bursa:2009we,DelDebbio:2010hx,DeGrand:2011qd,Catterall:2011zf,Bennett:2012ch}.
The most recent measurement obtained by fitting the mode number of the Dirac
operator gave a very precise value~\cite{Patella:2012da}. The mode number method has also been used to follow 
the running of $\gamma$ over a range of energy scales for the $SU(3)$ theory with many
light fundamental fermions~\cite{Cheng:2013eu}.

The large $N$ version of MWT, the $SU(N)$ gauge theory with two adjoint fermions, 
is interesting for several reasons.
From a phenomenological point of view, it is expected to be similar to
the $SU(2)$ theory. For example, the universal first two perturbative
coefficients of the beta function are independent of $N$.
Hence, as in
the case of MWT, they  point towards the existence of   an infrared fixed point with a mass anomalous 
dimension that is also independent of $N$.
 Moreover, numerical results for 
the mass anomalous dimension for the SU(2) and SU(3) theories appear to be in agreement~\cite{DeGrand:2013yja},
suggesting that this N--independence may be a good approximation all the way down to $N=2$.
From a more theoretical
point of view, the large $N$ theory is better suited for  connecting  with  results obtained from different 
approaches, such as the AdS/CFT correspondence. Fortunately, the  numerical study of the infinite volume 
theory at large $N$ is made possible by the concept of large $N$ volume
independence. This implies the equivalence with a single site lattice
reduced model, for which simulations can be performed at  large values
of $N$, that would be prohibitively expensive on a  conventional $L^4$ lattice. 
In this context, the study of large $N$ Yang-Mills theory with adjoint fermions has attracted much attention
~\cite{Kovtun:2007py}-~\cite{Bringoltz:2011by}. 
In this work we will be using the  twisted reduction technique~\cite{GonzalezArroyo:1982ub,GonzalezArroyo:1982hz}. 
For the adjoint fermion case the specific form of the action has been given in 
Ref.~\cite{Gonzalez-Arroyo:2013bta}. This model has been shown to lead to a 
softer $N$ dependence than the Adjoint Eguchi-Kawai model with periodic boundary 
conditions~\cite{Azeyanagi:2010ne,Gonzalez-Arroyo:2013bta}. The twisted
model depends on the choice of the twist tensor. Here we will follow the same
symmetric twist prescription as for the pure gauge theory, in which
$N$ is taken as the square of an integer number $N=\LL^2$, and the flux
through each plane is equal to $\pm k\LL$ (modulo $N$). With appropriate
values of the integer $k$, this  choice has proven effective 
in avoiding symmetry breaking for the pure gauge theory~\cite{GonzalezArroyo:2010ss}.
An important advantage of the twisted reduction method is that the dominant $1/N$
corrections amount to finite size effects on an $\LL^4$ lattice. This allows 
an estimate of the values of $N$ at which the simulations should be performed. 
In this work we will be using values of $N$ up to 289, corresponding to lattices 
of size $17^4$.

In summary, the purpose of this paper is to analyze the behaviour of the 
$SU(N)$ gauge theory with two flavours of adjoint fermions in the large
$N$ limit. Our main goal is to determine whether the theory has indeed a
non-trivial infrared fixed point (IRFP)  and to measure the mass anomalous
dimension at this fixed point. In previous papers some of the present 
authors studied the behaviour of Wilson loops and the corresponding
string tension~\cite{GonzalezArroyo:2012st,Gonzalez-Arroyo:2013dva,Gonzalez-Arroyo:2013yla,Gonzalez-Arroyo:2013gpa}.
Although the results were consistent with
the conformal behaviour characteristic of an IRFP, the extraction of the 
mass anomalous dimension had large uncertainties. Our methodology here 
will be based on an alternative procedure which has produced very
precise estimates in the study of MWT~\cite{Patella:2012da}. 
Preliminary results have been presented in Refs.~\cite{Keegan:2012xq}, \cite{Perez:2014gqa}.

The strategy is to determine the anomalous dimension from the structure 
of the eigenvalue density $\rho(\omega)$ of the massless 
Dirac operator $\slashed{D}$. The eigenvalue density is defined as 
\begin{equation}
\rho(\omega) = \lim_{V\rightarrow\infty} \frac{1}{V} \sum_k \delta(w -
w_k)\quad ,
\end{equation}
where the sum runs over all eigenvalues $i\omega_k$ of $\slashed{D}$.
In a mass--deformed conformal field theory (mCFT), this quantity 
should vanish for $w \rightarrow 0$ as~\cite{DelDebbio:2010ze}
\begin{equation}
\label{eq:rho}
\lim_{m\rightarrow 0}\lim_{V\rightarrow\infty}\rho(\omega) \propto
\omega^{\frac{3-\gamma_*}{1+\gamma_*}},
\end{equation}
where $\gamma_*$ is the mass anomalous dimension at the infrared fixed
point, $V$ is the lattice volume, and $m$ is the mass.  This behaviour should be 
contrasted with the one characteristic of a chirally broken theory where the eigenvalue density does not 
vanish at the origin.
 
In this paper we will use the previous idea to determine $\gamma_*$
for the large $N$ gauge theory with two adjoint quarks. It is clear 
from the previous formula that it is crucial to work in the region 
of very small masses and keeping finite volume effects under
control. As a reference  we will compare our result with those obtained 
for the pure gauge theory ($n_f=0$), for which the eigenvalue density 
has the characteristic behaviour of a chirally broken gauge theory.

The structure of the paper is as follows. In the next section we will 
collect all the technical aspects concerning the simulation and the 
extraction of the mass anomalous dimension from the data. In the
following we will present the results of our analysis. The
paper ends with the presentation of our conclusions.

\section{Methodology}

As explained in the introduction, our approach to the large $N$ limit is
based on reduction. Hence, we simulate the twisted reduced $SU(N)$ model 
on a single site with two adjoint Dirac  fermions~\cite{Gonzalez-Arroyo:2013bta}. 
In the large $N$ limit the theory is equivalent to the infinite volume
lattice gauge theory. For finite $N=\LL^2$, the corrections amount to
finite volume corrections in an $\LL^4$ lattice. Thus, it is important to 
keep track of the $N$-dependence which translates into the equivalent  
finite volume corrections. For that purpose we have performed
simulations at values of $N$ ranging from 16 up to 289, the latter 
corresponding to an effective lattice volume of  $17^4$. Our study has been
done at two values of $b$, 0.35 and 0.36, and a large number of
$\kappa$ values. The number of configurations used for the calculation of the eigenvalue spectrum
at each value of $b$, $\kappa$ and $N$ are listed in Tab.~\ref{tab:cnfgs}. 
In addition we calculated the eigenvalue spectrum
of the $n_f=0$ theory at $b=0.35,0.36$. For $N=121,289$, 
we used 10 configurations with $\kappa=0.170,0.175,0.180,0.185,0.190$, and for 
$N=841$, $b=0.36$ we used 4 configurations for $\kappa=0.190$.

\TABLE{
\begin{tabular}{c|c|c|c|c|c|c}
& & $N=16$ & $N=25$ & $N=49$ & $N=121$ & $N=289$ \\
\hline
& $k$ & 1 & 2 & 3 & 3 & 5 \\
& $\bar k / \sqrt{N}$ & 0.25 & 0.40 & 0.29 & 0.36 & 0.41 \\
\hline
\hline
b & $\kappa$ & $N=16$ & $N=25$ & $N=49$ & $N=121$ & $N=289$ \\
\hline
0.36 & 0.160 & 20k & 1200 & 500 & 20 & 20\\
 & 0.165 & - & - & - & 20 & 19\\
 & 0.170 & - & - & - & 20 & 15\\
\hline
0.35 & 0.160 & - & - & - & 40 & 20\\
 & 0.165 & - & - & - & 20 & 20\\
 & 0.170 & - & - & - & 20 & 13\\
\end{tabular} 
\caption{Number of configurations used for calculating the eigenvalue spectrum at each value of $b$, $\kappa$, $N$ and flux $k$. The integer parameter $\bar k$ satisfies $ k \bar k = 1$ (mod $\sqrt{N}$). 
All configurations are separated by at least 25 molecular dynamics updates, and at $N=289$ they are separated by 125 molecular dynamics updates. }
\label{tab:cnfgs}
}

As explained in the introduction, we
have chosen the symmetric twist configuration, with values of the
flux integer parameter $k$ given in Tab.~\ref{tab:cnfgs}.
These fulfill the condition $k/\sqrt{N} > 1/9$ which was found
necessary for the pure gauge theory (TEK model)  to respect 
the center symmetry~\cite{GonzalezArroyo:2010ss}. This symmetry is a
necessary  ingredient  in the proof of reduction by Eguchi and Kawai.
The addition  of light adjoint fermions should help in preserving the
symmetry but,  as shown in Ref.~\cite{Gonzalez-Arroyo:2013bta}, 
adhering to the condition allows the  study of the full range of $\kappa$
values and leads to a smoother $N$ dependence. As an example of the 
behaviour of Polyakov loops, which act as order parameters of the 
center symmetry, in Fig.~\ref{fig:poly} we display the expectation value
of the modulus of the unit winding loop as a function of $N$.  By
definition, this quantity is always positive, but as seen in the figure 
its size decreases with $N$ for all values of $\kappa$.

\FIGURE{
  \centering
    \includegraphics[angle=270,width=12.0cm]{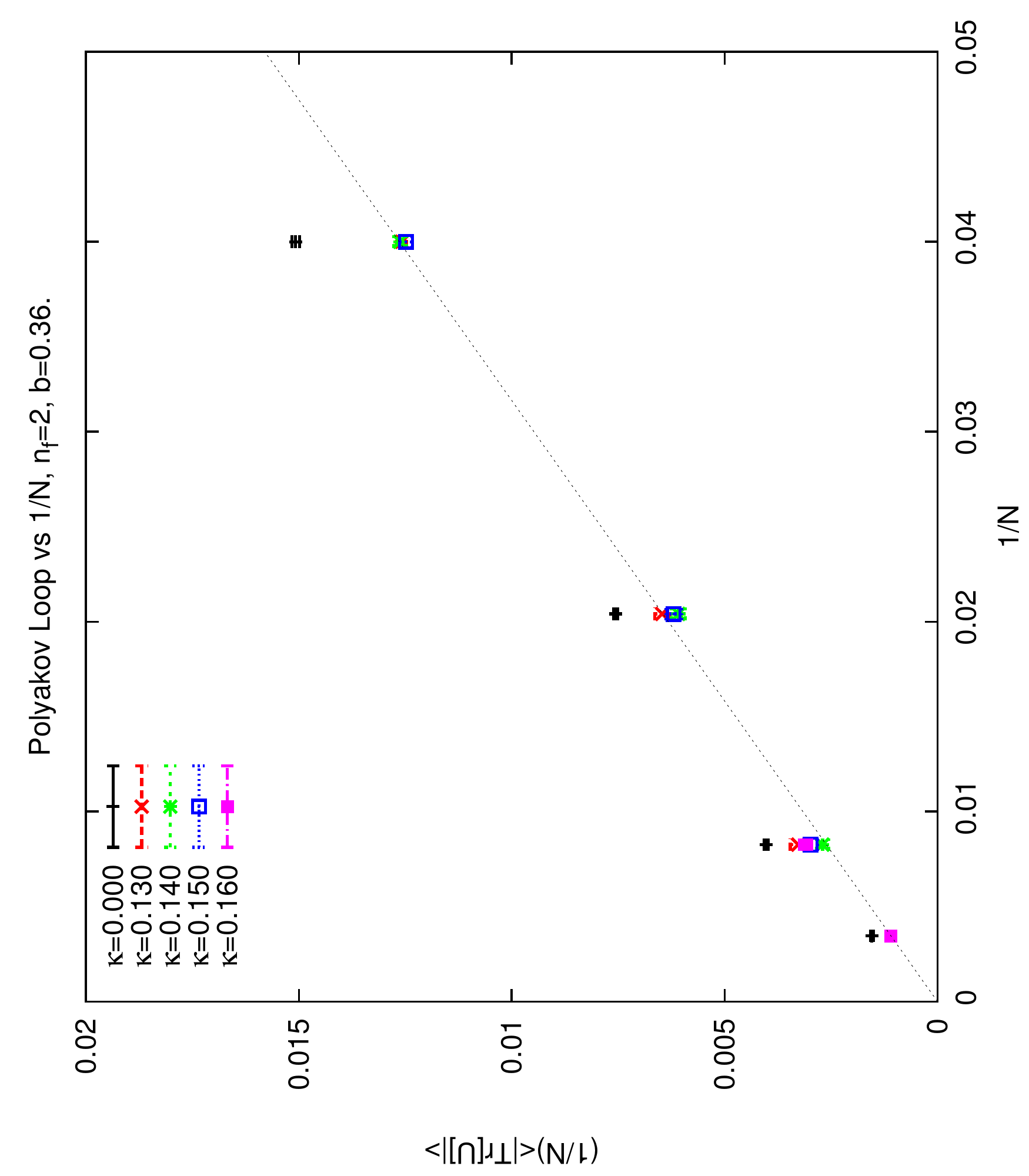}
  \caption{Polyakov loop: $\tfrac{1}{N}\left|\Tr U \right|$ vs $1/N$ for $n_f=2$. 
  Must go to zero in the large $N$ limit for reduction to hold, 
  which it does for all values of $\kappa$, where $\kappa=0$ corresponds to quenched $n_f=0$ data.}
  \label{fig:poly}
}

For each configuration we compute the low-lying spectrum of 
the modulus square of the massive lattice Wilson Dirac operator in the 
adjoint representation. This operator
is positive definite and in the naive continuum limit corresponds to 
$a^2(-\slashed{D}^2+m^2)$. Thus,  its eigenvalues,  labelled $(a\Omega)^2$,
are related to those of $\slashed{D}$ by the expression
\begin{equation}
a \omega=\sqrt{(a\Omega)^2-m^2a^2}\quad,
\end{equation}
where $a$ is the lattice spacing and 
$m$ is the adjoint quark mass.  
The lowest eigenvalue of our lattice operator defines the spectral gap. In the continuum it is bounded from below 
by $a^2m^2$. In the case of QCD and for the lattice Wilson Dirac operator, the median of the gap distribution was found empirically~\cite{DelDebbio:2005qa}  to satisfy the relation
\begin{equation}
a\Omega_0 \propto   \frac{1}{2\kappa} - \frac{1}{2\kappa_c}.
\end{equation}
so that $a\Omega_0$ is proportional to $am$. In our case, however, 
we expect that  the bound is not saturated at finite $N$. The reason 
being the absence of zero-momentum quark states in the reduced model. 
Hence, quarks are then produced with at least the minimum momentum 
$2\pi/L$, whose square decreases linearly with $1/N$. 

Since our main goal is the determination of the mass anomalous dimension
with the idea presented in the introduction, it is important to 
study the region close to the critical point and for large values 
of $N$ to minimize small effective volume corrections. For that reason,
our main  analysis was based on the study of  the lowest 2000 eigenvalues 
$(a\Omega)^2$ at $N=289$ and the lowest 1000 eigenvalues at $N=121$. The set of 
values of $b$ and $\kappa$ were given in Tab.~\ref{tab:cnfgs}. The
computational cost increases considerably as we approach $\kappa_c$,
explaining the smaller number of configurations for that case.
Fortunately, the distribution of the lowest lying spectrum does not
seem to fluctuate strongly at those values of $N$.

In determining the value of $\gamma_*$ from the distribution of
eigenvalues there are certain alternative procedures which we will
describe below.

\subsection{Determination of the mass anomalous dimension $\gamma_*$}

\subsubsection{Determining $\gamma_*$ from a fit to the spectral density}

 In the continuum $\gamma_*$ could be determined by fitting the spectral density to the form expected for a mass-deformed conformal field theory, Eq.~(\ref{eq:rho}).
However, in order to compare to the lattice data, it is more convenient to look at the spectral density of $-\slashed{D}^2$,  given by: 
\begin{equation}
\tilde \rho(\omega^2) = \lim_{V\rightarrow\infty} \frac{1}{V} \sum_k \delta(w^2 -
w_k^2) \propto
(\omega^2)^{\frac{1-\gamma_*}{1+\gamma_*}}.
\end{equation}
This quantity is obtained on the lattice by counting the number of eigenvalues of the modulus square of the Wilson Dirac operator 
within a bin of size $\Delta^2$ around $(a\Omega)^2$. Representing this number by ${\cal N}(a\Omega, \Delta)$, the lattice spectral density is given by:
 \begin{equation}
\tilde \rho_L((a\Omega)^2) = {1 \over N^2 \Delta^2}  {\cal N}(a\Omega, \Delta),
\end{equation}
where $N^2$ represents the lattice volume on the reduced lattice. 
The continuum formula for $\tilde \rho(\omega^2)$  gives the following parameterisation for the lattice data:
\begin{equation}
\label{eq:fitrho}
\tilde\rho_L((a\Omega)^2) = B\left[(a\Omega)^2-(am)^2\right]^{\frac{1-\gamma_*}{1+\gamma_*}},
\end{equation}
allowing the determination of the three free parameters: $B$, $(am)^2$ and $\gamma_*$. The lowest part of the eigenvalue distribution is 
the one most affected by finite volume and finite mass effects, hence the fits have to be performed in an intermediate range of eigenvalues
$a\Omega_{min}<a\Omega<a\Omega_{max}$,
which preserves  the separation of scales on the lattice,
\begin{equation}
\frac{1}{\sqrt{\mathrm{N}}} \ll am \ll  a\Omega  \ll 1.
\end{equation}
From a practical viewpoint the 3-parameter fit demands very precise data in a wide range of eigenvalues and induces strong correlations between the parameters. In some cases the number of parameters can be reduced by assuming that the mass is negligibly small ($am = 0$). Alternatively we can continue to work at finite mass but use information coming from the smallest eigenvalue to fix the parameter $(am)^2$ of the fit.

\subsubsection{Determining $\gamma_*$ from a fit to the mode number}

\FIGURE{
\centering
    \includegraphics[angle=270,width=12.0cm]{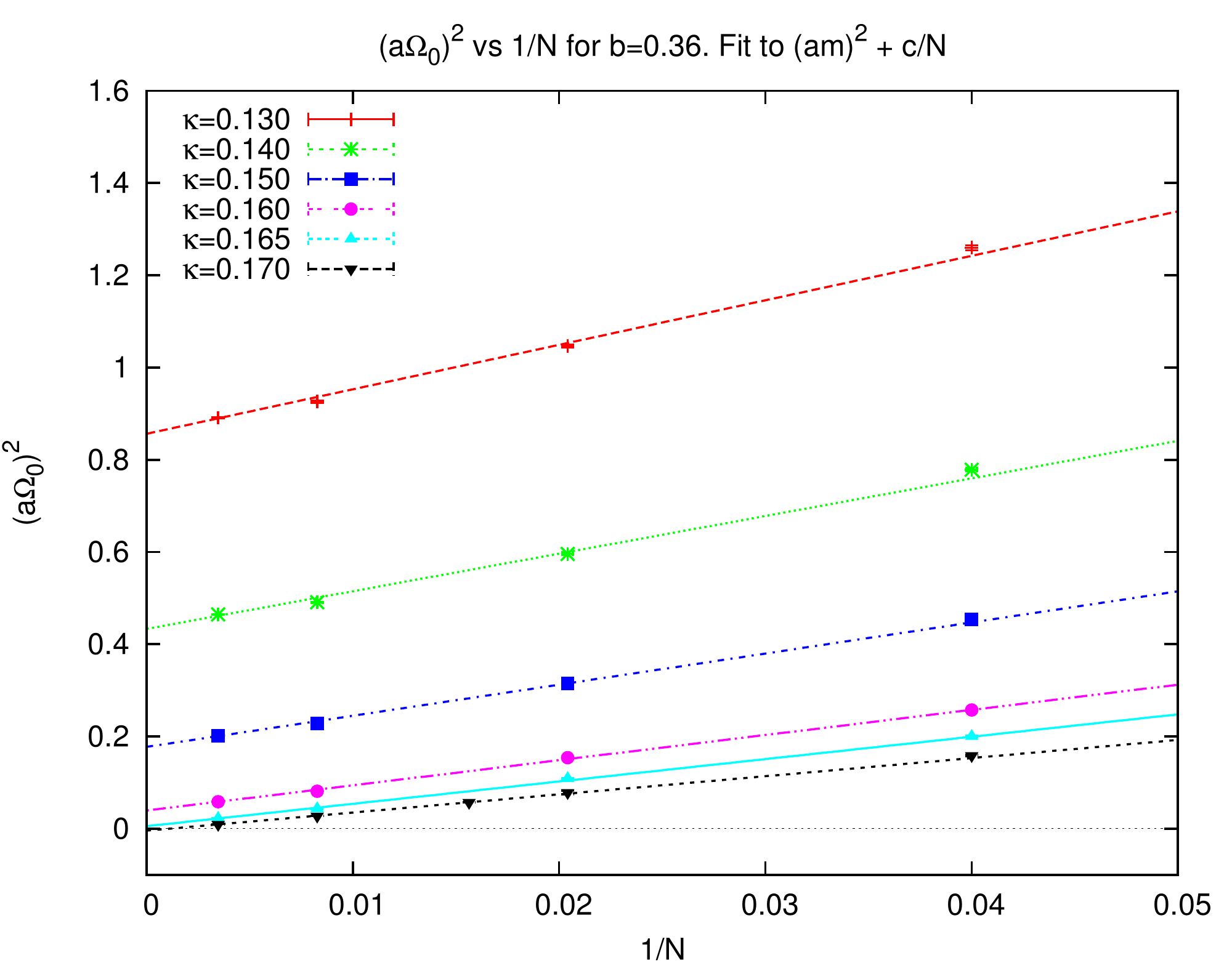}
  \caption{Median of the spectral gap of the modulus square of the Wilson Dirac operator, $({a\Omega}_0)^2$, vs $1/N$.
  The fitting form $({a\Omega}_0)^2 = (am)^2 + c/N$ fits all the data well.}
  \label{fig:msqvsN}
}

Alternatively we can follow the procedure introduced in Ref.~\cite{Patella:2012da} and extract $\gamma_*$ from the mode number 
${\nu}(\Omega)$ of the Dirac operator. It is simply defined as the number of eigenvalues of $(-\slashed{D}^2+m^2)$
below some value $\Omega^2$. Hence, it is given by $V$ times the integral 
of the eigenvalue density. We can split this integral into two 
parts as follows
\begin{equation}
{\nu}(\Omega) = 2 V \int_0^{\sqrt{\Omega_{IR}^2-m^2}}\rho(\omega)\,\, d\omega + 2 V \int_{\sqrt{\Omega_{IR}^2-m^2}}^{\sqrt{\Omega^2-m^2}}\rho(\omega)\,\, d\omega,
\end{equation}
The first part contains the range of eigenvalues which are more
sensitive to finite volume  and/or finite mass effects. For the second
part we can  insert Eq.~(\ref{eq:rho}) and perform the integration 
to give
\begin{equation}
\label{eq:fitIIIfull}
{\nu}(\Omega)  \simeq \nu(\Omega_{IR})-{\cal A} \left[\Omega^2_{IR}-m^2\right]^{\frac{2}{1+\gamma_*}}  +{\cal A} \left[\Omega^2-m^2\right]^{\frac{2}{1+\gamma_*}}.
\end{equation}
To determine ${\nu}(\Omega) $ on the lattice we just simply count the number of eigenvalues of the modulus square of the lattice Wilson massive Dirac operator below some value $(a \Omega)^2$. 
The continuum formula for the mode number implies that the lattice data 
can be parametrized as follows:
\begin{equation}
\nu_L(a\Omega) \simeq \nu_{0} +
A\left[(a\Omega)^2-(am)^2\right]^{\frac{2}{1+\gamma_*}},
\label{eq:fitI}
\end{equation}
where 
\begin{equation}
\nu_{0} = \nu_L(a\Omega_{IR}) -
A\left[(a\Omega_{IR})^2-(am)^2\right]^{\frac{2}{1+\gamma_*}}.
\end{equation} 
Eq.~(\ref{eq:fitI}), normalised dividing by the lattice volume,  is the expression used in Ref.~\cite{Patella:2012da}
to fit the lattice data. This allows the determination of its four free 
parameters ($\nu_{0}$, $A$, $(am)^2$ and $\gamma_*$). 

For the same reason as discussed before for the spectral density, one can attempt to reduce the number of parameters in the fit.
If we assume that finite volume (i.e. finite $N$) effects are negligible we might take $\Omega_{IR}$
close to threshold making $\nu_{0}$  negligibly small compared 
to the other term. This leads to a simplified expression 
\begin{equation}
\label{eq:fitIII}
\nu_L(a\Omega) \simeq A\left[(a\Omega)^2-(am)^2\right]^{\frac{2}{1+\gamma_*}}.
\end{equation}
which can be used to fit its three parameters ($A,am,\gamma_*$) to the
modenumber data in a range of eigenvalues $\Omega \in [\Omega_{min},\Omega_{max}]$
 satisfying:
\begin{equation}
\frac{1}{\sqrt{\mathrm{N}}} \ll am \ll a\Omega_{IR} < a\Omega <
a\Omega_{UV} \ll 1.
\label{eq:scales}
\end{equation}

If we set $am=0$ in Eq.~(\ref{eq:fitIII}) we can reduce the number 
of free parameters even further. This is for example the strategy adopted in Ref.~\cite{Cheng:2013eu} 
where a 2--parameter fit to the mode number is used,
\begin{equation}
\label{eq:fitII}
\nu_L(\Omega) \simeq
A[(a\Omega)^2]^{\frac{2}{1+\gamma_*}}.
\end{equation}
The sensitivity to the fit function, the volume, the mass parameter, or the fitting range will be used to estimate the systematic error in the 
determination of the mass anomalous dimension.

\TABLE{
\begin{tabular}{c|c|c|c|c}
$b$ & $\kappa$ & $N=121$ & $N=289$ & $N=\infty$ \\
\hline
0.36 & 0.160 & 0.0803(6) & 0.0585(3) & 0.0429(7) \\
 & 0.165 & 0.0433(4) & 0.0224(2) & 0.0074(5) \\
 & 0.170 & 0.0279(4) & 0.0096(2) & -0.0036(4) \\
\hline
0.35 & 0.160 & 0.0997(9) & 0.0815(4) & 0.0683(9) \\
 & 0.165 & 0.0530(5) & 0.0346(2) & 0.0214(6) \\
 & 0.170 & 0.0281(4) & 0.0105(4) & -0.0021(7) \\
\end{tabular} 
\caption{Lowest eigenvalues squared for each $b$, $\kappa$ and $N$, along with the extrapolation to $N=\infty$ which corresponds to the mass parameter $(am)^2$.}
\label{tab:mass}
}

\section{Results}

\subsection{Analysis of the two flavor case}
\FIGURE{
\centering
    \includegraphics[angle=0,width=\linewidth]{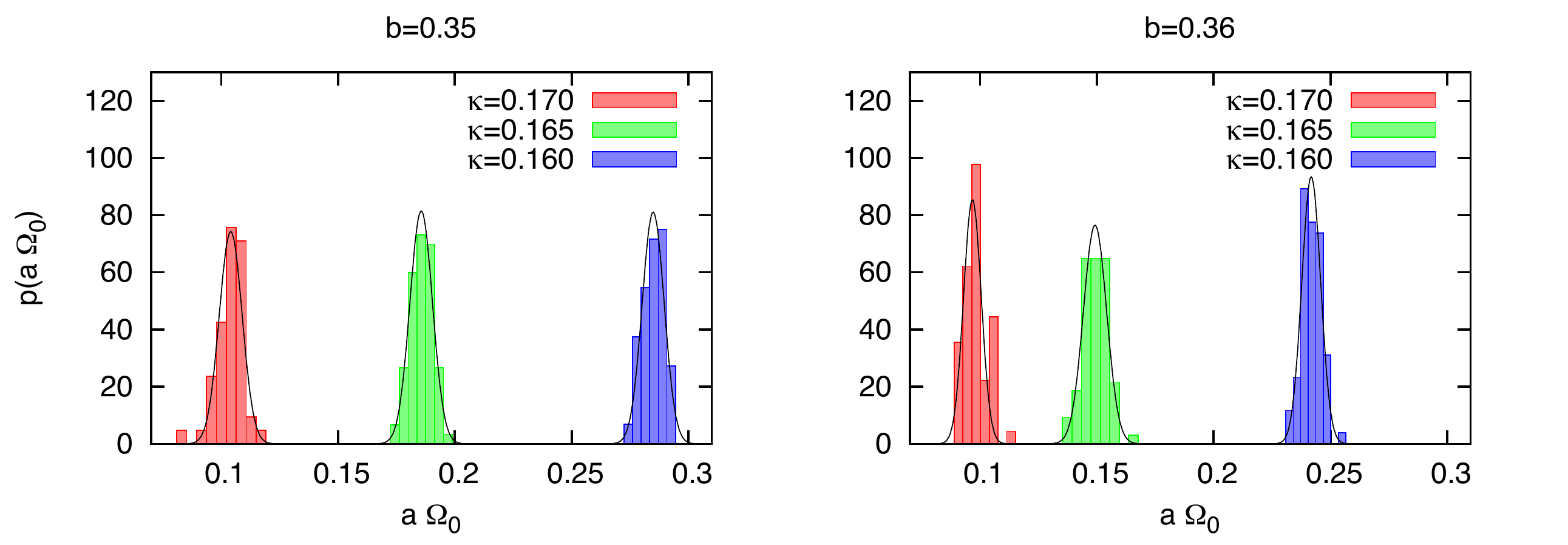}
  \caption{Probability distribution of the spectral gap of the Wilson Dirac operator for  $N=289$ and various values of $\kappa$.
The lines are fits to a gaussian distribution, Eq.~(\ref{eq:gapdist}).}
  \label{fig:gapdist}
}

\FIGURE{
\centering
    \includegraphics[angle=0,width=0.8\linewidth]{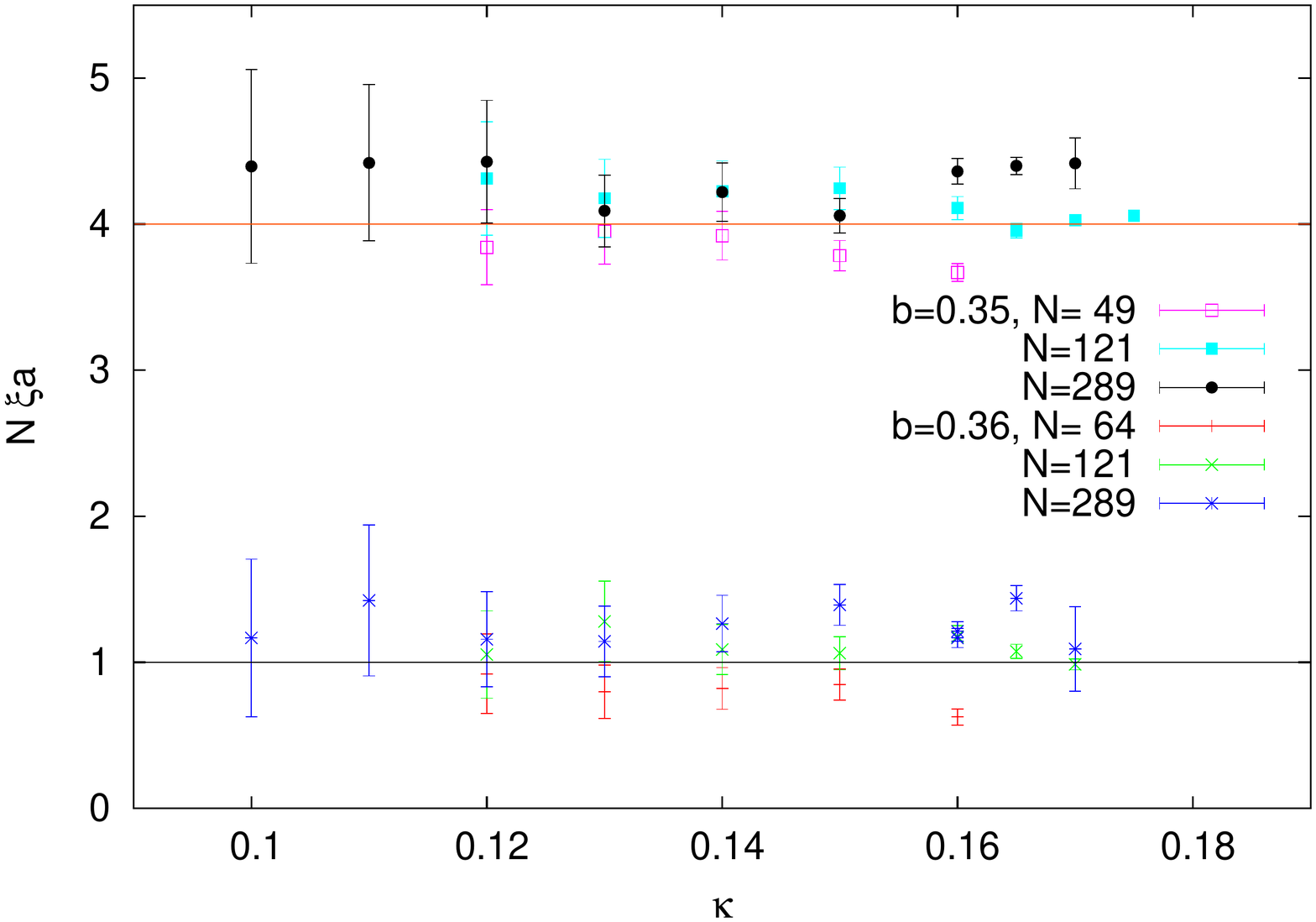}
  \caption{We display $N \xi a$ for our ensemble of $n_f=2$ configurations for $b=0.36$ and $b=0.35$ (displaced by 3 vertically for clarity), where $\xi a$ is the width  of the probability distribution of the spectral gap of the Wilson Dirac operator. }
  \label{fig:sigmagap}
}

In this section we will present the results of our study. 
Our first step is the analysis of  the spectral gap of the
hermitian Wilson--Dirac operator. To study how this 
quantity behaves as a function of $N$ and $\kappa$,
we measure the lowest eigenvalue at $b=0.36$ on a range of configurations 
for $N=25-289$ and $\kappa=0.130-0.170$.
As argued in the previous section, in the twisted model we expect the median of the gap distribution
$({a \Omega}_0)^2$ to differ from $(am)^2$ by a finite volume 
correction which, interpreted as non-zero momentum contribution, should 
depend linearly on $1/N$. Indeed, a linear fit of this kind seems to
describe our non--perturbative data  quite well, as shown in
Fig.~\ref{fig:msqvsN}.  This allows us to determine the 
mass parameter $(am)^2$ in the large $N$ limit, listed in Tab.~\ref{tab:mass}.

\FIGURE{
\centering
    \includegraphics[angle=270,width=12.0cm]{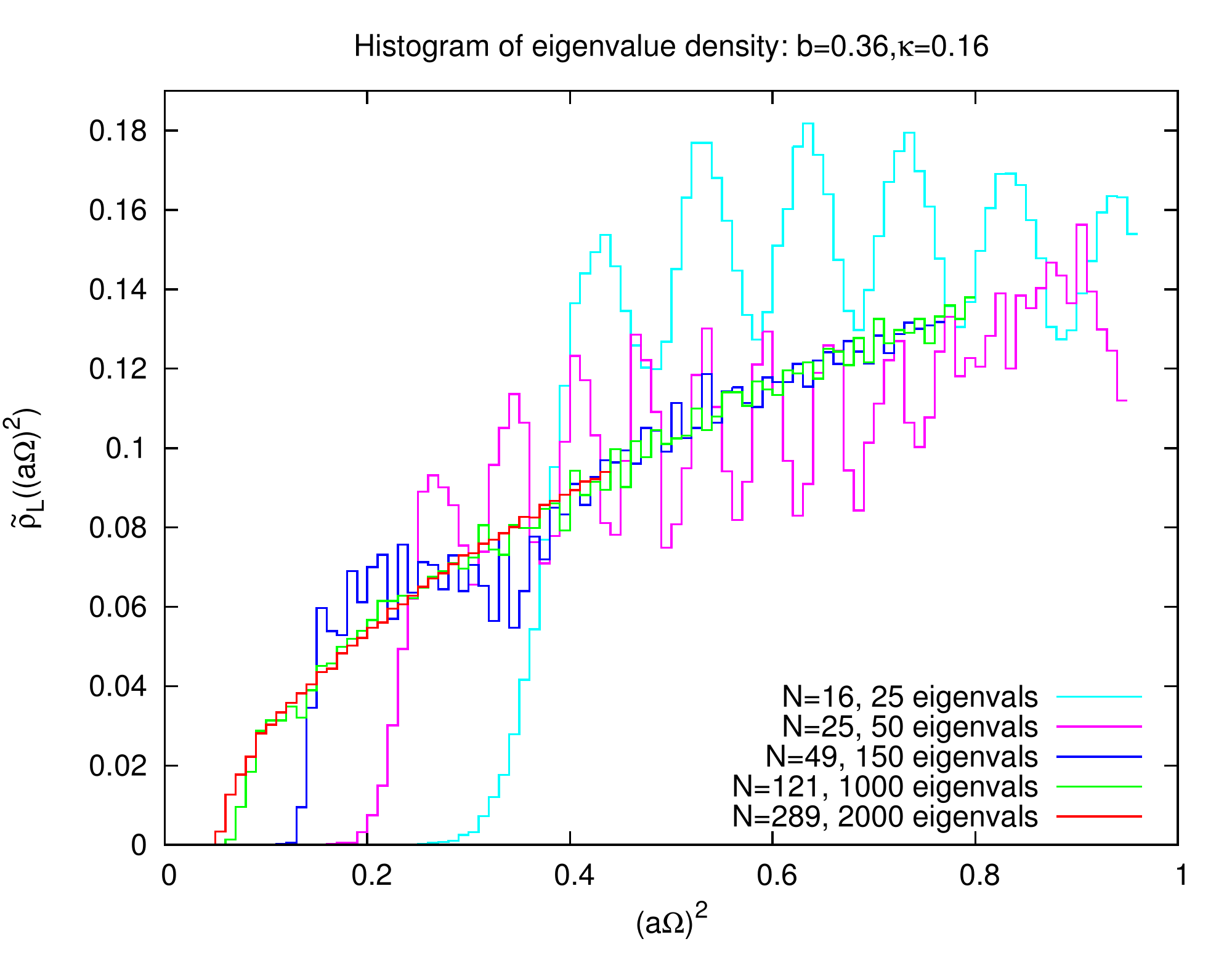}
      \caption{Spectral density $\tilde\rho_L((a\Omega)^2)$ for $b=0.36$,
      $\kappa=0.160$,
        with a bin--size $\Delta^2=0.01$. There is agreement at all
        eigenvalues (except
          the very lowest couple of bins) between $N=121$ and $N=289$. For
          $N=49$ there is a
            qualitative difference for the lower eigenvalues, but the
            large ones are also in agreement.}
              \label{fig:histogram_b36}
              }

We have analysed the probability distribution of the spectral gap $p({a \Omega}_0)$ of the Wilson Dirac operator. In the large volume limit of QCD, 
the analysis in Ref. \cite{DelDebbio:2005qa} showed that the distribution is gaussian:
\be
p({a \Omega}_0) \propto  \exp \Big \{ - {1 \over 2 (\xi a)^2} \left( a \Omega_0 - \left\langle {a \Omega}_0 \right\rangle \right)^2\Big \}\,
\label{eq:gapdist}
\ee
with median proportional to the bare current quark mass and width scaling with the volume and the lattice spacing approximately as
$\xi a = a^2 / \sqrt{V}$. In the reduced lattice this would imply $\xi a= 1/N$. Fig.~\ref{fig:gapdist} shows $p({a \Omega}_0)$ for $N=289$ at $b=0.35$ and $0.36$ and
several values of $\kappa$. A gaussian fit describes the data well. The fitted distribution widths multiplied by $N$ are displayed in Fig.~\ref{fig:sigmagap} for our $n_f=2$ configurations at several values of $N$ and $\kappa$. Our results follow rather well the behaviour also reported in QCD.

Let us now move on to  describe our results for the distribution of
eigenvalues. One of the main points is to analyze the  $N$-dependence 
of this distribution. We already saw that this dependence affects the 
gap of the spectrum, but we expect this effect to have a small impact 
for higher eigenvalues. 
This can be seen in Fig.~\ref{fig:histogram_b36}, which shows a histogram of the number of
eigenvalues as a function of $(a\Omega)^2$, for $b=0.36$, $\kappa=0.160$.
Comparing different values of $N$, we see agreement between $N=289$ and $N=121$ in all but the 
first two bins. For $N=49$, for $(a\Omega)^2 \lesssim 0.4$ the behaviour is qualitatively 
different, but for higher eigenvalues we again see agreement with larger values of $N$. For 
$N=25$ and $N=16$ there are strong oscillations in the distributions,
which are presumably the sum of the distributions of individual eigenvalues with the allowed
discrete momenta.

\FIGURE{       \centering
\includegraphics[angle=270,width=7cm]{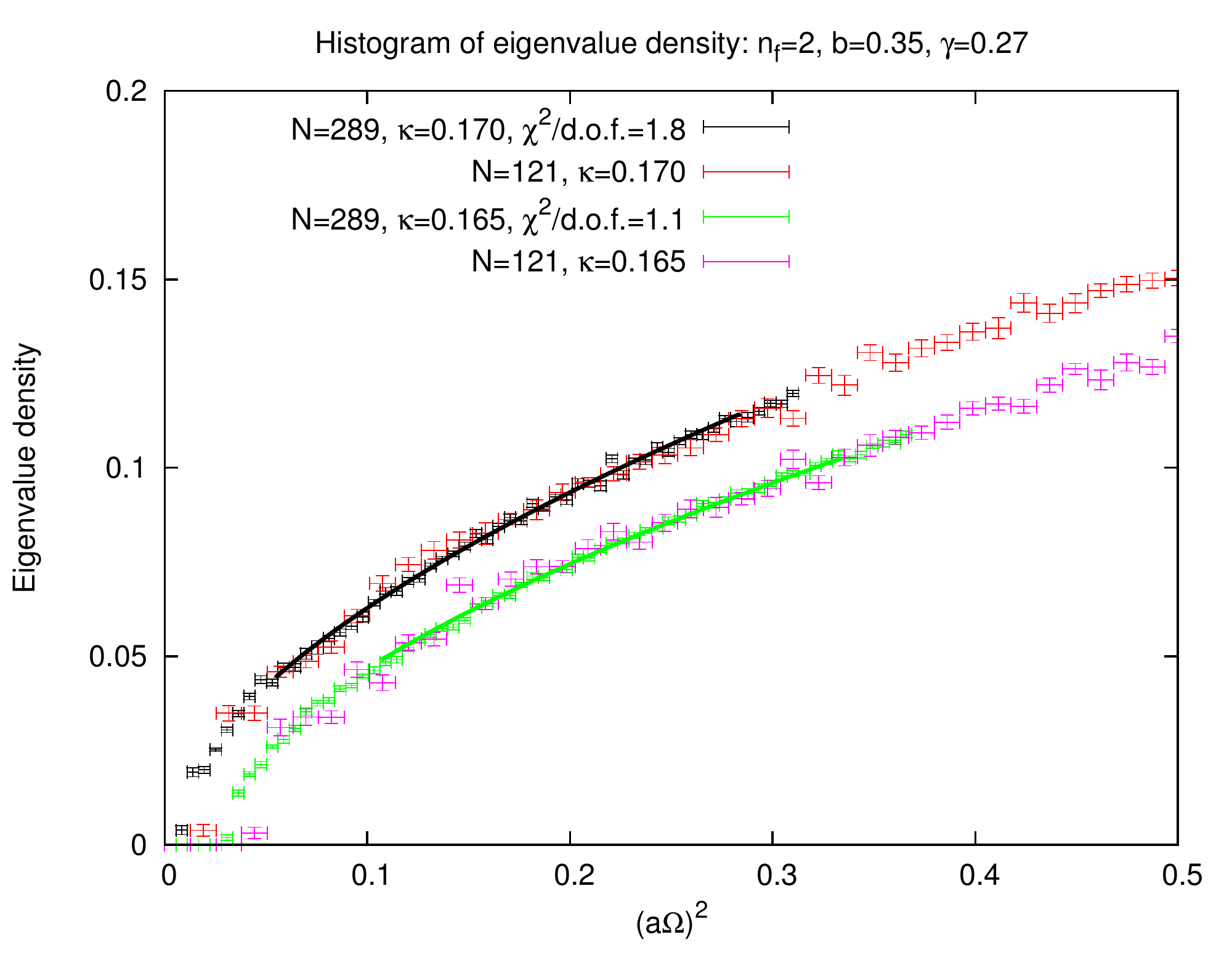}
\includegraphics[angle=270,width=7cm]{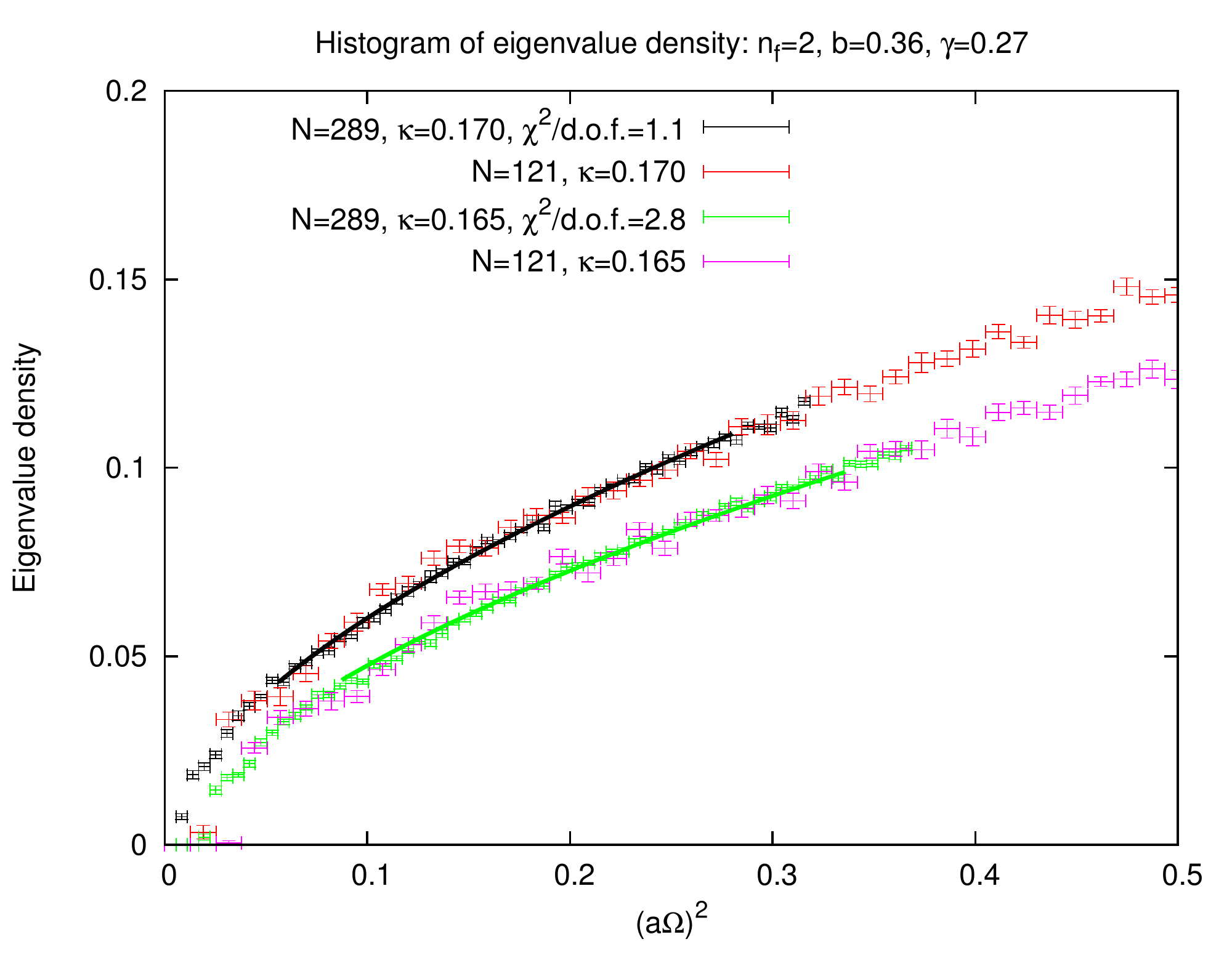}
     \captionof{figure}{Eigenvalue density distribution at $b=0.35$ (left) and $b=0.36$ (right)
      for $\kappa=0.165,0.17$ and $N=121,289$.}
        \label{fits}
          }

We have seen that, at least for the two biggest values of $N$, 
the eigenvalue distribution roughly coincides beyond a certain 
threshold value. The question now is to see if this distribution
behaves as expected from the IRFP hypothesis and to extract $\gamma_*$
from it. In the previous section we gave two alternative methods of
fitting the data. One is to compare the eigenvalue distribution with
        Eq.~(\ref{eq:fitrho}). The other  is to compute the mode
	number and
	fit it to  Eq.~(\ref{eq:fitIIIfull})  or to its simplified expression
	Eq.~(\ref{eq:fitIII}).

 In performing a fit one has to select the range of values 
 ${\Omega}_{\mathrm{min}} < \Omega < {\Omega}_{\mathrm{max}}$ 
 to be fitted. A lower value of ${\Omega}_{\mathrm{min}}$ 
 increases the sensitivity to the value of the mass $(am)^2$
 but also risks to be more affected by finite effective volume (finite
 $N$) corrections. For the mode number fit the same is expected to 
 happen for the parameter $\nu_0$. Furthermore, the narrower
 the fitting range the stronger the correlations among parameters 
 leading to high uncertainties in $\gamma_*$.

 If we use only the data which is least affected by finite volume and
 finite mass effects, we can produce our most precise determination of
 $\gamma_*$: Hence, we take our data of $N=289$ and $\kappa=0.17$ 
 and fit the distribution of eigenvalues to Eq.~(\ref{eq:fitrho}) 
 with $(am)^2$ set to zero. The upper edge of the fitting range
 $(a\Omega_{\mathrm{max}})^2$ covers almost all of the $N=289$ data. 
 For the lower edge $(a\Omega_{\mathrm{min}})^2$ the cut is set to
 twice the lowest eigenvalue for $N=121$. The result for $b=0.36$ 
 is $\gamma_*=0.268(2)$ and for $b=0.35$ is $\gamma_*=0.271(1)$. 
 It is remarkable that both values of $b$ give consistent results 
 within the purely statistical 1\% errors. To give an idea of 
 the quality of the fit we display the data in
 Fig.~\ref{fits} together with the best fit. In the
 figure we also include the data
 of $N=121$ at both values of $b$. The fitted
 function  also describes well the behaviour of the data at $N=121$,
 except for the smallest eigenvalues where finite volume effects should 
 be mostly felt. The continuous line going through the $\kappa=0.165$
 data was obtained fitting the corresponding data with $\gamma_*$ fixed 
 to the value obtained at $\kappa=0.17$ and the mass square $(am)^2$
 to the large $N$ extrapolated value given in Tab.~\ref{tab:mass}.

              \FIGURE{       \centering
              \includegraphics[angle=270,width=12cm]{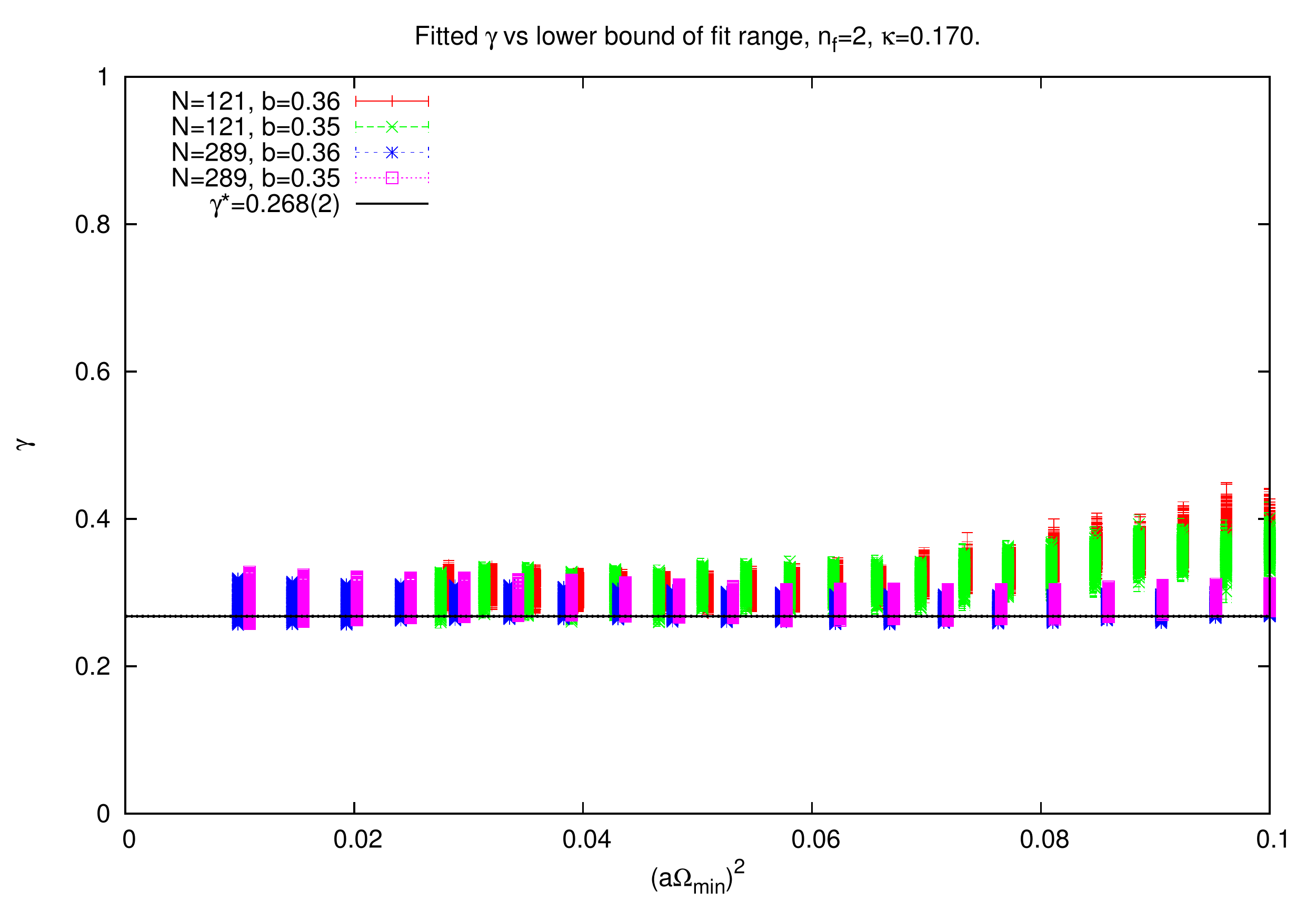}
\caption{Fitted $\gamma_*$ as a function of
$(a\Omega_{\rm min})^2$ from fitting the spectral density
data to Eq.~(\ref{eq:fitrho}).
For each point, a
range of masses $(am)^2$ between zero and the
lowest eigenvalue at $N=289$ are used, and the
upper fit range bound $(a\Omega_{\rm max})^2$ is also
varied over a range of values from 0.25
up to the largest value for which we have data.
The black line shows our best estimate for $\gamma_*$
with statistical error only.}
\label{fig:fitsyst}
          }

The analysis of the previous figure and fits
 indicates that our data look consistent with the predictions of an infrared fixed point with 
 mass anomalous dimension close to $0.27$. 
 In order to substantiate the claim of conformality, it is important to check the systematic uncertainties involved in our analysis and to compare
 our results with those corresponding to theories which are not conformal in the IR. 
 The latter will be done in the next section where we will compare our $n_f=2$ results with those obtained for $n_f=0$. 
 What we will now present is an evaluation of the systematic errors. This point will be analysed by estimating
 the effect of finite $N$ corrections, finite mass corrections and 
 sensitivity to the fitting range. We will also test what results come 
 out if we use the mode number instead of the eigenvalue density.  
 
 A good summary of the effect of all systematics on the value of
 $\gamma_*$ is provided by Fig.~\ref{fig:fitsyst}. Here we display
 different determinations of $\gamma_*$ for the $N=289$ and $N=121$ data by varying 
 $\Omega_{\max}$, $\Omega_{\min}$ and $(am)^2$. The latter is varied 
 within the range extending from zero to the minimum $N=289$ eigenvalue. 
 The data is displayed as a function of $(a\Omega_{\min})^2$. For 
 a large range of x-axis values all determinations of $\gamma_*$
 (obtained by using  different values of $\Omega_{\max}$ and  $(am)^2$) 
 fall within a horizontal 
 strip whose width serves as an upper bound to our systematic error
 $\delta \gamma_*= 0.05$. For larger values of $(a\Omega_{\min})^2$ 
 the fitting range narrows, obviously leading to a wider spread of 
 values of $\gamma_*$.

 To better analyze  the dependence of  the fitted value of $\gamma_*$ with the 
 remaining parameters, we display in Fig.~\ref{fig:fit_mdep} its 
 correlation with  the mass parameter used in the fits for all 
 values  $(a\Omega_{\max})^2$ and $(a\Omega_{\min})^2<0.07$.
 This shows that the main source of systematic errors is indeed the value of 
 the mass. 

\FIGURE{
  \centering
      \includegraphics[angle=270,width=0.9\linewidth]{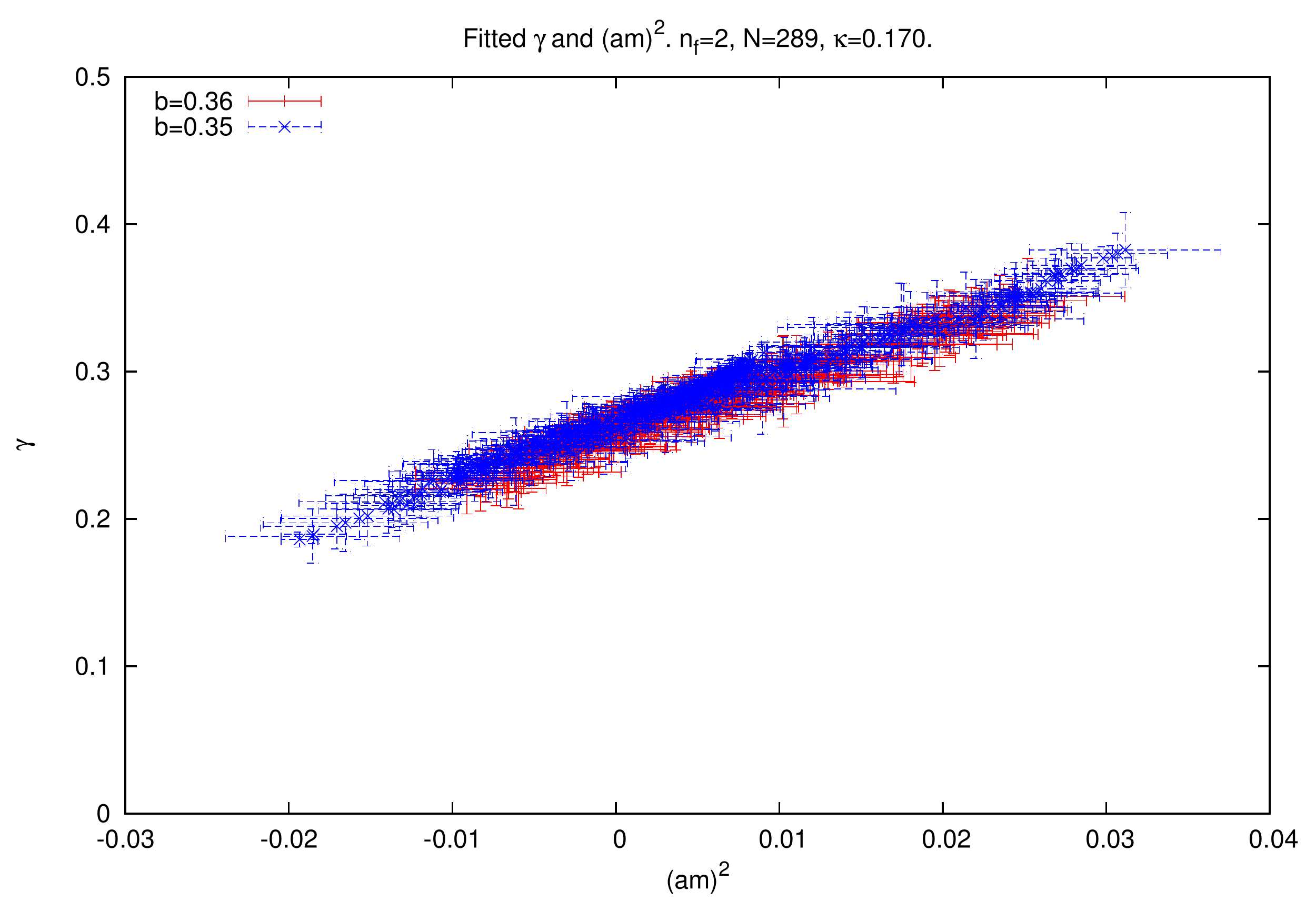}
        \caption{Fitted values of $(am)^2$ and $\gamma_*$ when both are
	taken as free parameters when fitting the spectral 
	  density data to Eq.~(\ref{eq:fitrho}), using $N=289$
	  configurations. There is a clear correlation between the two 
	    parameters.}
	      \label{fig:fit_mdep}
	      }

\FIGURE{       \centering
\includegraphics[angle=270,width=7.4cm]{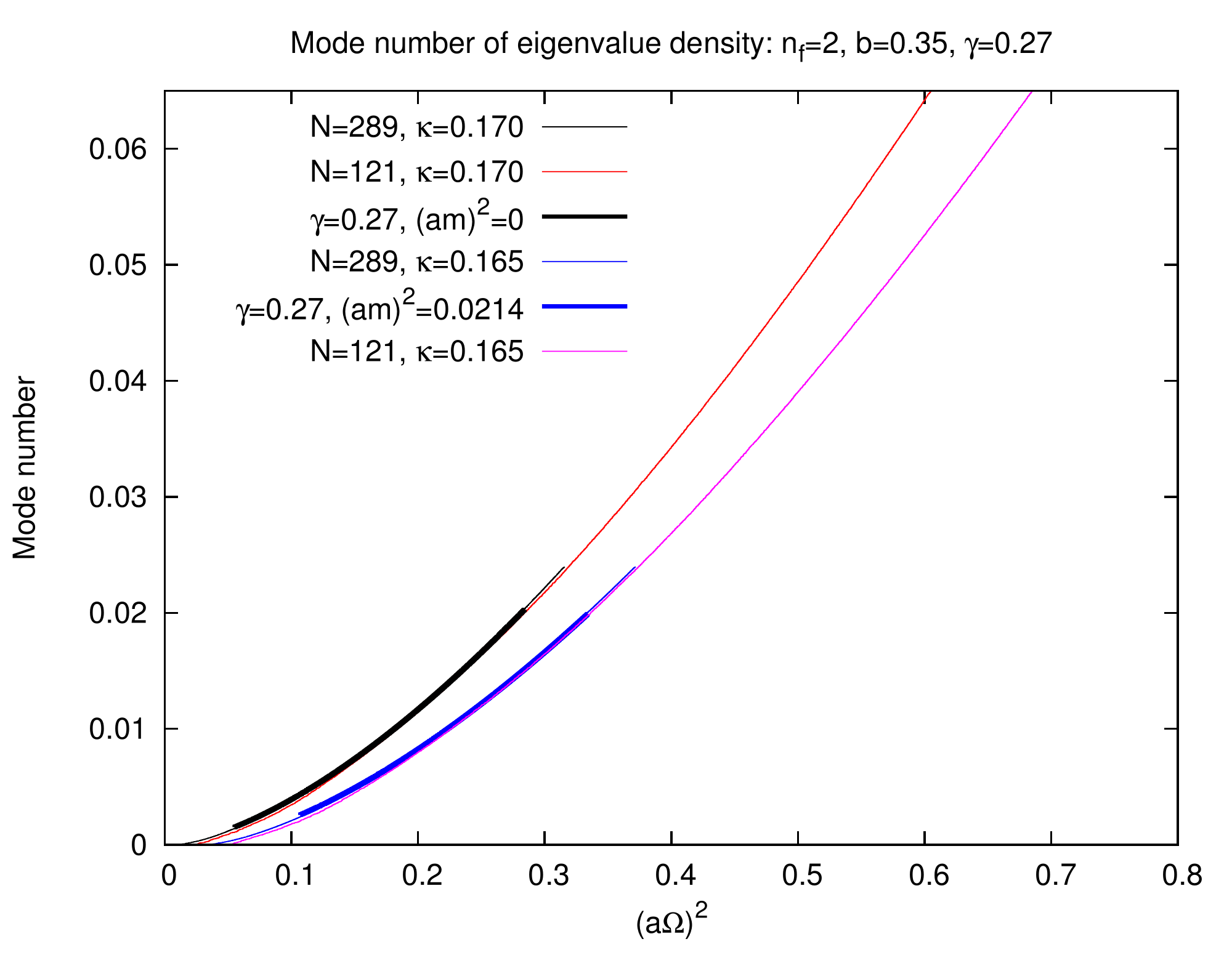}
\includegraphics[angle=270,width=7.4cm]{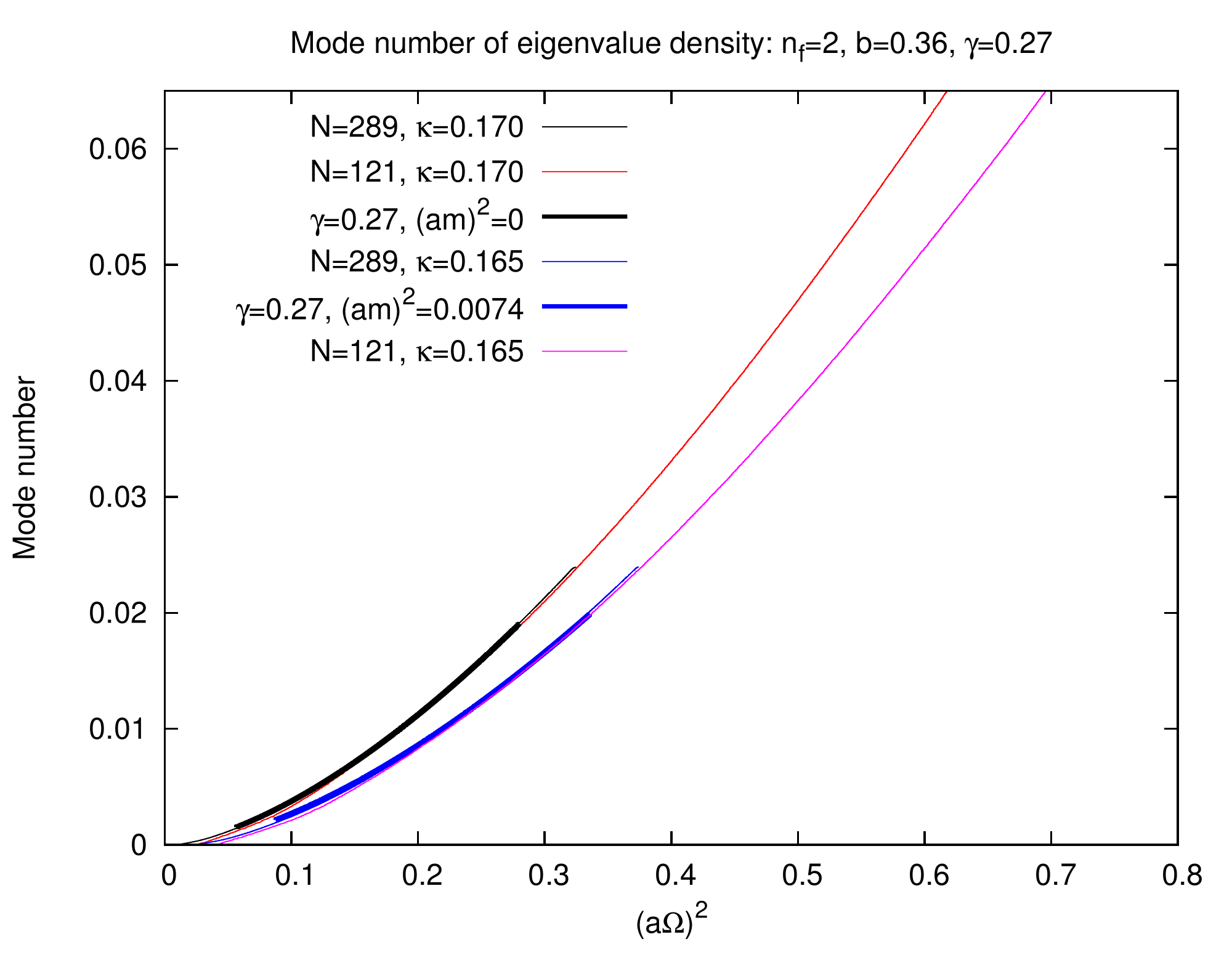}
     \captionof{figure}{Fits to the mode number at $b=0.35$ (left) and $b=0.36$ (right)
      for $\kappa=0.165,0.17$ and $N=121,289$.}
        \label{fig:mode}
          }

 \FIGURE{
  \centering
        \includegraphics[angle=0,width=\linewidth]{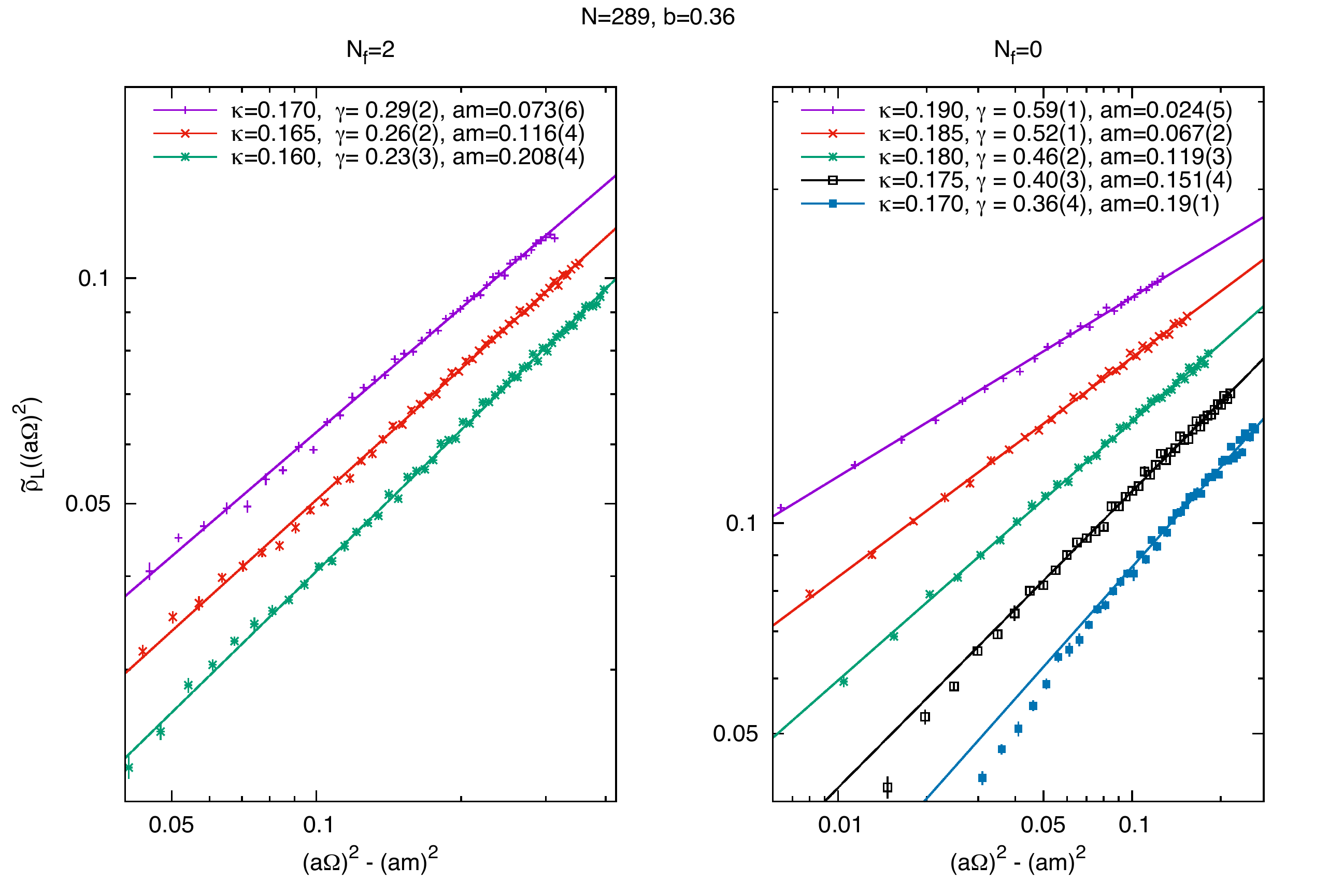}
                \caption{Eigenvalue density distribution, in log-log scale, at $b=0.36$ for
$n_f=2$  and $n_f=0$.
The lines correspond to the fits to Eq. ( \ref{eq:fitrho}) described in the text.
}
\label{fig:compare0_nf}
           }

A similar analysis can be done for the mode number. Results are
essentially compatible. In this case, one has an issue about which 
fitting formula one should use. Although the 4-parameter formula
Eq.~(\ref{eq:fitIIIfull}) should apply in all regions, there
are strong correlations between the fitted parameters, so that for example a
large range of values of $\gamma_*$ can produce a good fit by a suitable choice of 
the other 3 fit parameters.
The more restrictive fitting formulas better constrain the fitted quantities, 
but have a limited range of applicability. For example Fig.~\ref{fig:mode} shows a fit to Eq.~(\ref{eq:fitIII})
using the same configurations and fit ranges as for the eigenvalue density in Fig.~\ref{fits}, which
gives $\gamma_*=0.270(2)$ for $b=0.36$ and $\gamma_*=0.272(1)$ for $b=0.35$, which are 
in good agreement with the numbers determined from the eigenvalue density.

\subsection{Comparison between zero and two flavors}

The comparison between our $n_f=2$ results with those obtained for $n_f=0$ is essential to substantiate the claim of conformality. The
latter theory is not conformal and should display a different
behaviour. Indeed, for the pure Yang-Mills theory it makes no sense to speak
of $\gamma_*$ itself, since there is no infrared fixed point. However,
it still makes sense to study the behaviour of the spectral density
and mode number distribution as a function of its argument.

To carry on the previous study, we generated pure gauge configurations at the same
values of $b$ and $N$ and a range of $\kappa$ values, and repeated our
previous analysis for this data. The  $\kappa$ values were chosen in
such a way as to explore similar small values  of the minimum eigenvalue.

\FIGURE{
\centering
\includegraphics[angle=0,width=\linewidth]{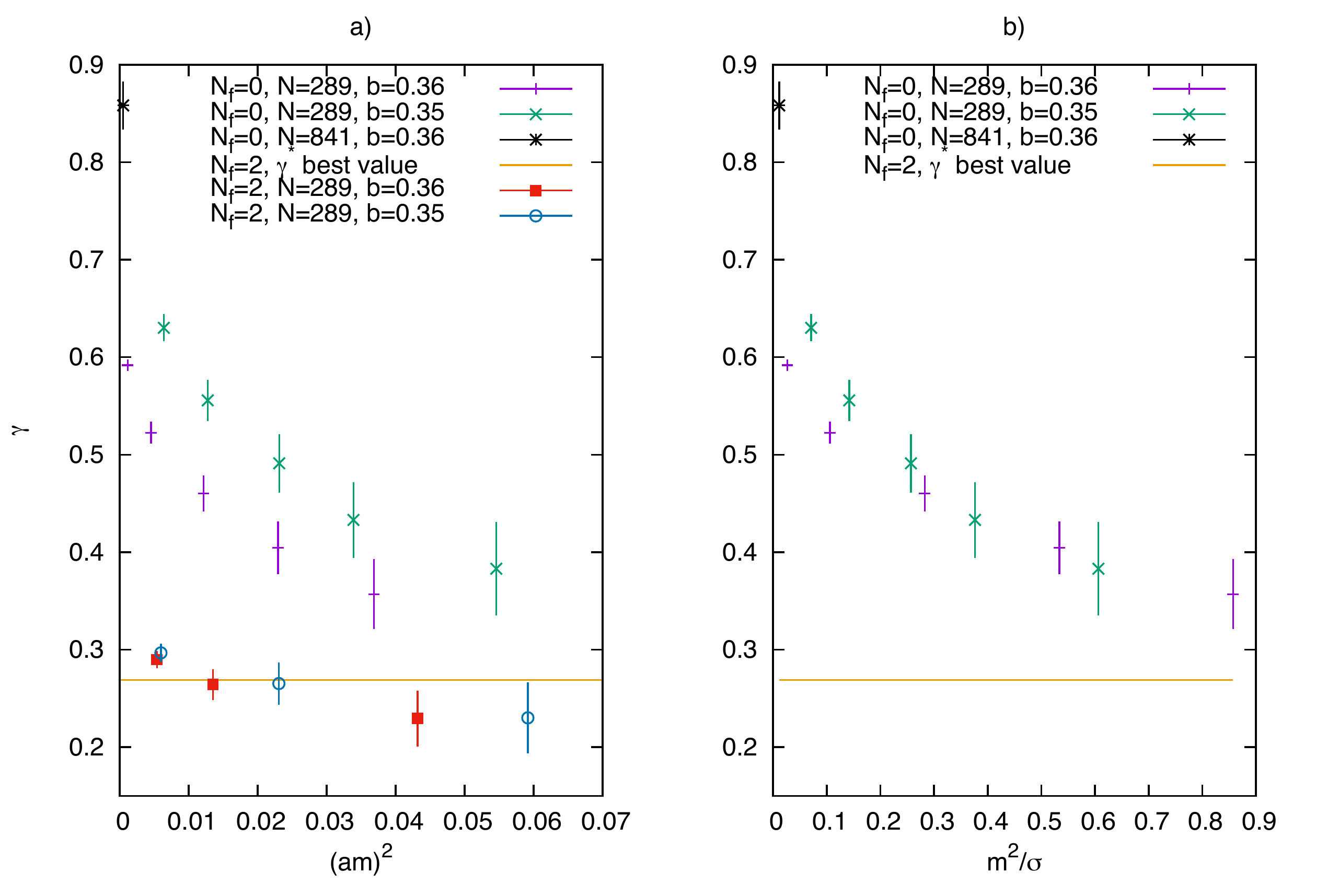}
\caption{a) Dependence of the extracted value of $\gamma_*$  with the mass $(am)^2$ for $n_f\!=0\!$ and
      $n_f\!=\!2$. b) Dependence of the extracted value of $\gamma_*$  with $(m)^2/\sigma$ for $n_f\!=0\!$.
  }
 \label{fig:compare2_nf}
}

The fits to the spectral density for the largest
values of $\kappa$ were qualitatively as good as those of $n_f=2$.
Fig. \ref{fig:compare0_nf} shows, in log-log scale, the fits to the lattice spectral densities for
$n_f=2$  and $n_f=0$ at $b=0.36$.
The values of $\gamma_*$ in this plot are obtained by
fitting the spectral densities to Eq. ( \ref{eq:fitrho}).
The mass $(am)^2$ is varied in the range $0.5(a\Omega_0)^2 \leq (am)^2 \leq (a\Omega_0)^2$, the number of
bins is varied from 40 to 80, and the lower edge of the fit range is varied in the range $1.5(a\Omega_0)^2 \leq a\Omega_{\rm min} \leq 2(a\Omega_0)^2$.
The final point is an average over all of these choices for many bootstrap replicas of the data.
Keeping $\gamma_*$ fixed to this average, the fits are repeated for each set of data in order to determine $am$.
For this final fit the lower edge of the fit range is set to $2(a\Omega_0)^2$.
In all cases the $\chi^2$ per degree of freedom of the fit is below 2.
Hence, from the quality of the fit to the spectral density one cannot deduce the presence of an infrared fixed point.

One remarkable difference appears when looking at the dependence of the extracted value of $\gamma_*$  with the mass $(am)^2$.
The result is shown in Fig. \ref{fig:compare2_nf}a.
The value of $\gamma_*$ seems quite stable for the $n_f=2$ case. This is what one expects in the vicinity of an
infrared fixed point as the anomalous dimensions tends to a constant at the fixed point. The result for $n_f=0$ is quite different, showing
a pronounced drop as we move away from the critical value of $\kappa$. For the smallest masses the fitted value of $\gamma_*$ reaches as high
values as $0.8$. Notice that a value equal to $3$ would imply a constant value of the spectral density at the origin.
Our data show a growing $\gamma_*$ for lighter masses 
but do not yet reach the value of 3 predicted by chiral symmetry breaking, through the Banks-Casher formula.
This is probably due to finite volume and/or finite mass effects.

Another marked difference between the two cases appears in the dependence of the spectral densities on the bare inverse coupling $b$.
In the conformal case one expects the coupling to be a marginally irrelevant operator close to the IRFP,
in contrast to the $n_f=0$ case where it is marginally relevant and determines the lattice spacing.
The values of $\gamma_*$ displayed in Fig. \ref{fig:compare2_nf}a for $n_f=2$ do indeed have a very small dependence on $b$.
However, this dependence is large for the case of $n_f=0$.
In the quenched case, the results corresponding to the two different couplings only show scaling when expressed in physical units in terms
of the string tension. This is shown in Fig. \ref{fig:compare2_nf}b, where we have used the data for the $n_f=0$ string tension obtained in
Refs. \cite{GonzalezArroyo:2012fx}
($\sigma a^2 = 0.09$ and $\sigma a^2 = 0.043$ for $b=0.35$ and $b=0.36$ respectively).

\section{Conclusions}

We have performed a measurement of the mass anomalous dimension $\gamma_*$
of the SU($N$) gauge theory with  two adjoint Dirac fermions, in the large $N$ limit
using the concept of large $N$ twisted reduction. Results from a   single site
lattice model at large values of $N$, have the expected qualitative behaviour
of the spectral density and mode number of the adjoint massless Dirac
operator. The distribution for small masses  (extracted from the
lowest eigenvalue) can be well-fitted with the expectations of an IRFP.
From the data we extract a value of $\gamma_*=0.269\pm 0.002 \pm
0.05$, where the first error is statistical and the second one
systematic.  This value is similar to  previous
lattice determinations of this quantity for the SU(2) theory.

Does our result provide conclusive evidence of the presence of an
infrared fixed point for the SU($\infty$) gauge theory with 2 flavours
of adjoint fermions? To try to answer this question we repeated the
analysis for the $n_f=0$ case, which is known to have a
completely different behaviour at criticality. However,  we observe
that the spectral densities at fixed $b$ and small quark mass,
can also be fitted with the same formulas with  a larger value
of $\gamma_*$. This conveys a word of warning about  drawing
conclusions about the existence of an IRFP only  from the capacity
to fit the spectral density or mode number  with a powerlike distribution.
Nevertheless, there are marked differences between the behaviour
observed in the  $n_f=0$ and $n_f=2$ cases. One of them is the
dependence of the result on the bare coupling $b$. The $n_f=2$ results
for our two values of $b$, 0.35 and 0.36, are consistent with each
other. This is not the case for the $n_f=0$ data. Although
insensitivity to the bare coupling $b$ is certainly the expected
result for an IRFP, it is difficult to exclude the fact that this is
not simply due to the smaller value of the  beta function when
adding fermions in the adjoint. The second difference refers to the
change of behaviour as  we approach criticality $\kappa\longrightarrow
\kappa_c$. The extracted value of $\gamma_*$ for $n_f=2$ remains
fairly stable as one expects if the behaviour is indeed dictated by the
presence of an IRFP. On the contrary for the $n_f=0$ data we observe a
pronounced  rise of the value of $\gamma_*$ as we approach criticality.

To improve on these results using this method, smaller fermion masses and
larger volumes (i.e. larger values of $N$) would be required. Since finite
volume effects are the dominant limitation, one possibility for future work
would be to extend the single site lattice to a $2^4$ lattice.

\appendix
\section*{Acknowledgments}
We acknowledge financial support from the MCINN 
grants FPA2012-31686 and FPA2012-31880, 
and the Spanish MINECO's ``Centro
de Excelencia Severo Ochoa'' Programme under grant
SEV-2012-0249. M. O. is supported by the Japanese MEXT grant No
26400249 and the MEXT program for promoting the enhancement of research universities. 
Calculations have been done on Hitachi SR16000 supercomputer
both at High Energy Accelerator Research Organization(KEK) and YITP in
Kyoto University, and the HPC-clusters
at IFT. Work at KEK is supported by the Large Scale Simulation Program
No.14/15-03.

\end{document}